\documentclass[preprint,12pt]{elsarticle}

\usepackage{amssymb}

\begin{document}

\begin{frontmatter}



\title{Simulation of Ultra-High Energy Photon Propagation with PRESHOWER 2.0}

\author[a1]{P. Homola\corref{cor1}}
\ead{Piotr.Homola@ifj.edu.pl}
\author[a2]{R. Engel}
\author[a1]{A. Pysz}
\author[a1]{H. Wilczy\'nski}
\cortext[cor1]{Corresponding author: Tel.: +48 12 6628348; fax: +48 12 6628012.}
\address[a1]{H.~Niewodnicza\'nski Institute of Nuclear Physics, Polish Academy of Sciences, ul.~Radzikowskiego 152, 31-342 Krak\'ow, Poland}
\address[a2]{Karlsruhe Institute of Technology, 76021 Karlsruhe, Germany}
\begin{abstract}
In this paper we describe a new release of the PRESHOWER program, a tool for Monte Carlo simulation of propagation of ultra-high energy photons in the magnetic field of the Earth. The PRESHOWER program is designed to calculate magnetic pair production and bremsstrahlung and should be used together with other programs to simulate extensive air showers induced by photons.
The main new features of the PRESHOWER code include a much faster algorithm applied in the procedures of simulating the processes of gamma conversion and bremsstrahlung, update of the geomagnetic field model, and a minor correction. The new simulation procedure increases the flexibility of the code so that it can also be applied to other magnetic field configurations such as, for example, encountered in the vicinity of the sun or neutron stars.

\end{abstract}


\end{frontmatter}

\section{Program Summary}

\begin{tabular}{p{7cm}p{7cm}}
Program title: & PRESHOWER 2.0\\
Catalog identifier: & ADWG\_v2\_0\\
Program summary URL: & http://cpc.cs.qub.ac.uk/summaries/ ADWG\_v2\_0.html\\
Program obtainable from: & CPC Program Library, Queen's University, Belfast, N. Ireland\\
Licensing provisions: & Standard CPC licence, http://cpc.cs.qub.ac.uk/licence/licence.html\\
Programming language: & C, FORTRAN 77 \\
Computer(s) for which the program has been designed: & Intel-Pentium based PC \\
Operating system(s) for which the program has been designed: & Linux or Unix\\
RAM required to execute with typical data: & $<100$ kB \\
CPC Library Classification: & 1.1. \\
External routines/libraries used: & IGRF~\cite{igrf-paper,tsygan}, DBSKA~\cite{cernbess}, ran2~\cite{numrec} \\
Catalog identifier of previous version: & ADWG\_v1\_0 \\
Journal Reference of previous version: & Computer Physics Communications 173 (2005) 71-90 \\
Does the new version supercede the previous version?: & yes \\
Nature of problem: & Simulation of a cascade of particles initiated by UHE photon in magnetic field. \\
Solution method: & The primary photon is tracked until its conversion into an $e^+e^-$ pair. If conversion occurs each individual particle in the resultant preshower is checked for either bremsstrahlung radiation (electrons) or secondary gamma conversion (photons).\\
Reasons for the new version: & 1) Slow and outdated algorithm in the old version (a significant speed up is possible); 2) Extension of the program to allow simulations also for extraterrestrial magnetic field configurations (e.g.\ neutron stars) and very long path lengths. \\
\end{tabular}

\begin{tabular}{p{7cm}p{7cm}}
Summary of revisions: & A veto algorithm was introduced in the gamma conversion and bremsstrahlung tracking procedures. The length of the tracking step is now variable along the track and depends on the probability of the process expected to occur. The new algorithm reduces significantly the number of tracking steps and speeds up the execution of the program. The geomagnetic field model has been updated to IGRF-11, allowing for interpolations up to the year 2015. Numerical Recipes procedures to calculate modified Bessel functions have been replaced with an open source CERN routine DBSKA. One minor bug has been fixed.\\
Restrictions: & Gamma conversion into particles other than an electron pair is not considered. Spatial structure of the cascade is neglected.\\
Running time: & 100 preshower events with primary energy $10^{20}$~eV require a 2.66 GHz
CPU time of about 200 sec.; at the energy of $10^{21}$~eV, 600 sec.\\

\end{tabular}

\newpage
\section{Introduction}

Identifying and understanding the sources of cosmic rays with energies up to $10^{20}$\,eV is one of the most important questions in astroparticle physics 
(see, for example, \cite{Bluemer:2009zf,LetessierSelvon:2011dy,Kotera:2011cp}). Knowing the fraction of photons in the flux of ultra-high energy cosmic 
rays is of particular importance as photons are unique messengers of particular source processes (acceleration vs. decay of super-heavy particles or 
other objects). They are also produced in interactions of charged cosmic ray nuclei of the highest energies with cosmic microwave background radiation, 
known as the Greisen-Zatsepin-Kuzmin (GZK) effect. If the energy of the cosmic ray particles exceeds the GZK energy threshold at the source, ultra-high 
energy photons are produced due to well-understood hadronic interactions with microwave photons and can be detected at Earth as a unique propagation 
signature. So far only upper limits to the photon flux at ultra-high energy exist, which have led to severe constraints on models for UHECR sources.
However, depending on the primary cosmic ray composition, the sensitivity of the latest generation of cosmic ray detectors, i.e.\ the Pierre Auger
 Observatory~\cite{auger} and the Telescope Array~\cite{Tokuno:2010zz}, should allow detection of GZK photons for the first time.

The simulation of the propagation of photons before they reach the Earth's atmosphere is important because of the preshower 
effect \cite{preshw-mcbreen} that may occur when a photon traverses a region where the geomagnetic field component transverse 
to the photon trajectory is particularly strong. As described e.g.\ in Refs.~\cite{cpc1,phot-rev}, high energy photons in the 
presence of a magnetic field may convert into e$^{+}$e$^{-}$ pairs and the newly created leptons emit bremsstrahlung photons, 
which again may convert into e$^{+}$e$^{-}$ if their energies are high enough. As a result of these interactions, instead of a 
single high energy photon, a shower of particles of lower energies, the so-called preshower, reaches the atmosphere. The occurrence of 
the preshower effect has a large impact on the subsequent extensive air shower development and changes the predicted shower observables.

In 2005 the program PRESHOWER~\cite{cpc1} for simulating the showering of photons in the Earth's magnetic field was released. 
It was shown that this initial version of the Monte Carlo code is in good general agreement with previous 
studies \cite{preshw-mcbreen,karakula,stanev,billoir,bedn,vankov1,vankov2}.
In this paper we describe the changes of the PRESHOWER code relative to the initial version 1.0 \cite{cpc1}. 
The main feature of the new release is a much faster algorithm for calculating the distance at which a preshower interaction (gamma conversion or bremsstrahlung) occurs. In version 1.0, the calculations were done in constant steps along the particle trajectory and the step size was optimized for all possible trajectories. The constant step practically disabled studying the preshower effect along paths longer than several tens of thousands kilometers. Now, in version 2.0, the distance to the next interaction point is computed with an efficient veto algorithm, decreasing significantly the number of computing steps. The new method, being independent of the trajectory length, allows computations for arbitrarily long photon paths, e.g.\ simulations of preshower creation in the vicinity of a neutron star or active galactic nucleus. The new algorithm is described in detail in Section~\ref{sec1} of this article. Other important changes, i.e.\ the update of the geomagnetic field model and a code correction in version 1.0 are discussed in Section~\ref{sec2}. The results of testing PRESHOWER~2.0 are presented in Section~\ref{sec3} and conclusions are given in Section~\ref{sec4}.

Following Ref.~\cite{cpc1}, all the results presented in the following are obtained for the magnetic conditions of the Pierre Auger Observatory in Malarg\"ue, Argentina (35.2$^\circ$S, 69.2$^\circ$W). The shower trajectories are given in the local frame where the azimuth increases in the counter-clockwise direction and $\phi=0^\circ$ refers to a shower coming from the geographical North.

\section{The new sampling algorithm}
\label{sec1}

In PRESHOWER 1.0 the effect of precascading is simulated following the particle trajectories with a fixed step size. In each step the probability of conversion into e$^{+}$e$^{-}$ is calculated for photons and the probability of emitting a bremsstrahlung photon is computed for electrons. The step size has to be optimized for all possible trajectories and magnetic field configurations encountered along the particle trajectories. It has been found out that a step size of $10$\,km works well until the primary photon conversion and then the emission of bremsstrahlung photons as well as conversions of secondary photons are simulated in steps of $1$\,km. In this algorithm a typical simulation run consists of several thousands of steps.

The number of simulation steps and hence the computing time can be significantly reduced by using a veto algorithm. The algorithm used in the following is commonly applied in physical situations where a probability of occurring of a certain process varies within a given spatial or temporal interval (see, e.g.\ Ref.~\cite{veto}). The location or time of the occurrence of a physical process studied is found in few approximating jumps. An example of a process that can be treated with the algorithm is a radioactive decay considered in a certain interval of time. Photon conversion and bremsstrahlung probabilities to be found along spatial trajectories can also be computed with this veto algorithm.

\subsection{General description}
\label{algorithm}

The theory behind the veto algorithm is based on a process of discrete events described by
\begin{equation}
\frac{dN}{dt} = - f(t)\, N(t).
\end{equation}
Then the probability $dP_A$ of occurrence of process $A$ in the time window $t \dots t +dt$ is given by
\begin{equation}
d P_A  = - \frac{1}{N(t)} \frac{dN}{dt} = f(t) dt\ .
\end{equation}
Together with the probability of not having an occurrence of process $A$ in the time from $t_0$ to $t$
\begin{equation}
P_{{\rm no-}A} = \frac{N(t)}{N(t_0)}
\end{equation}
one obtains for the probability $dP$ for having an occurrence of $A$ in the time window $t \dots t +dt$, provided that this process did not occur at an earlier time $t^\prime$ with $t_0 < t'<t$
\begin{equation}
\label{eq1}
 dP = P_{{\rm no-}A}\, d P_A = f(t)\, \frac{N(t)}{N(t_0)} \,dt = f(t)\, \exp\left\{-\int^t_{t_0} f(t')\,dt'\right\}\, dt\ .
\end{equation}
If an analytic solution can be found for the integral of $f(t)$
\begin{equation}
F(t) = \int_{t_0}^t f(t^\prime)\, dt^\prime
\end{equation}
one can sample the time $t$ of the next occurrence of $A$ after the previous occurrence at time $t_0$ using the inversion method
\begin{equation}
\label{eq4}
 \int^t_{t_0} dP = \exp\{F(t)\} = \xi,
 \hspace*{2cm}
t=F^{-1}(\ln \xi),
\label{eq5}
\end{equation}
with $\xi$ being a random number uniformly distributed in $(0,1]$.

If $F(t)$ cannot be found or the inverse of it computed sufficiently easily one can use a function $g(t)$ such that $\forall t \geq 0: g(t) \geq f(t)$ and apply the following procedure
\begin{enumerate}
\item set the initial conditions: $i=0$, $t_0=0$;
\item $i=i+1$;
\item get a random number $\xi_i\in(0,1)$;
\item compute $t_i=G^{-1}(G(t_{i-1})-\ln \xi_i)$, $t_i>t_{i-1}$;
\item get another random number $\xi_i'\in(0,1)$;
\item if $\xi_i'\leq f(t_i)/g(t_i)$ then $t_i$ is the wanted result, i.e.\ the moment when process $A$ occurred, 
otherwise one has to go back to step 2 or, if the end of the interval is reached, end the procedure without occurrence of process $A$.
\end{enumerate}

The algorithm described above was mathematically proven to reproduce exactly the expected distributions~\cite{veto-proof}.

\subsection{Implementation in PRESHOWER 2.0}
\label{veto_presh}

Following the general scheme described above, the application of the veto algorithm to simulations of the preshower effect is straightforward. 
Instead of the time variable $t$, the distance $r$ along the preshower trajectory is used. 
We consider two processes in parallel over an interval starting at $r_{start}$ and ending at $r_{end}$, namely gamma conversion and 
bremsstrahlung of electrons. Following the physics notation introduced in Ref.~\cite{cpc1} (see also the Appendices A and B for all the required 
physics formulas and symbols) the probability functions are defined as
\begin{equation}
p_{conv}(r) \equiv \alpha(\chi(r))
\label{pconv}
\end{equation}
(see Eqs.~\ref{npairs}-\ref{pconv2}) for gamma conversion and
\begin{equation}
p_{brem}(r) \equiv \int^E_0 I(B_\bot(r),E,h\nu)\frac{d(h\nu)}{h\nu}
\label{pbrem}
\end{equation}
(see Eqs.~\ref{daug}-\ref{bremprob}) for magnetic bremsstrahlung. The function $f(t)$ is then replaced by $p_{conv}(r)$ or $p_{brem}(r)$, depending on the process to be simulated.
Since finding the antiderivatives of $p_{conv}(r)$ or $p_{brem}(r)$ is not straightforward, simple functions limiting $p_{conv}(r)$ and $p_{brem}(r)$ are used. We define
\begin{equation}
\begin{array}{c}
 g_{conv}(r) \equiv p^{max}_{conv}=const,\: \forall r \in (r_{start},r_{end}): p^{max}_{conv} \geq p_{conv}(r),\\
 g_{brem}(r) \equiv p^{max}_{brem}=const,\: \forall r \in (r_{start},r_{end}): p^{max}_{brem} \geq p_{brem}(r),
\end{array}
\end{equation}
which replace $g(t)$ and for which the antiderivatives are
\begin{equation}
\begin{array}{c}
 G_{conv}(r) = p^{max}_{conv} \cdot r,\\
 G_{brem}(r) = p^{max}_{brem} \cdot r.
\end{array}
\end{equation}
With these substitutions the algorithm of Sec.~\ref{algorithm} is applied in PRESHOWER 2.0 for the two interaction processes.

The determination of $p^{max}_{conv/brem}$ is crucial for computing time optimization. 
Too large $p^{max}_{conv/brem}$ increases the total number of steps to be executed in the procedure.

\subsection{Determination of $p^{max}_{conv/brem}$}

\begin{figure}
\begin{center}
\includegraphics[width=1.0\textwidth]{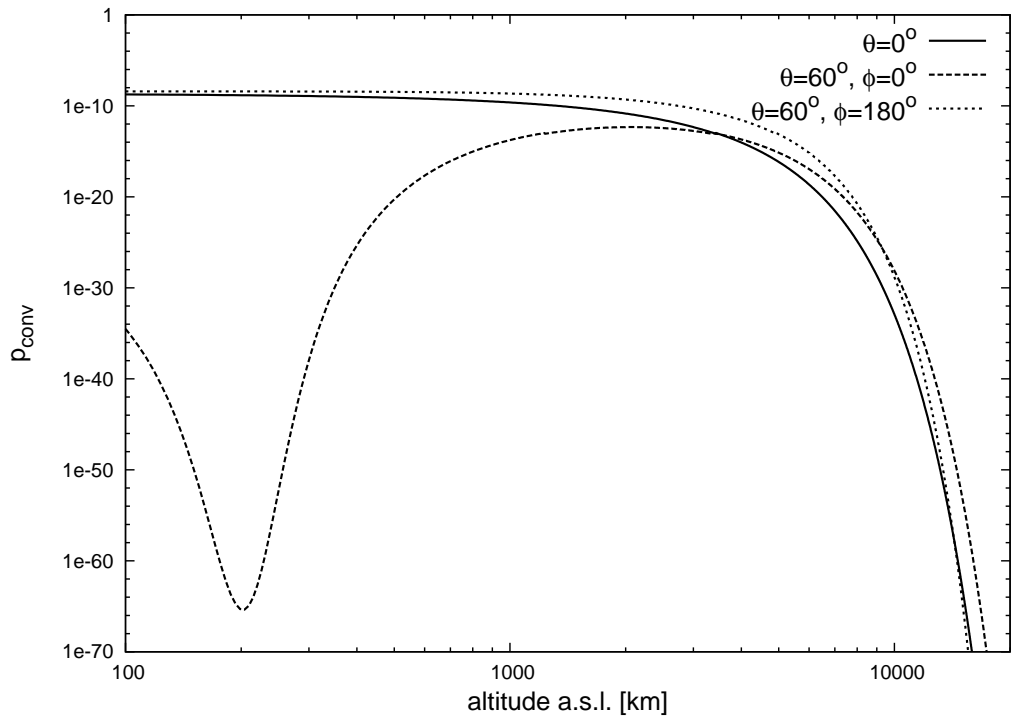}
\end{center}
\caption {
Examples of $p_{conv}$ functions along different trajectories at the location of the Pierre Auger Observatory in  Malarg\"ue (Argentina). A minimum value for one of the curves is related to the small value of $B_\bot$ for this specific arrival direction and altitude. See text for further details.}
\label{alpha_conv}
\end{figure}

\begin{figure}
\begin{center}
\includegraphics[width=1.0\textwidth]{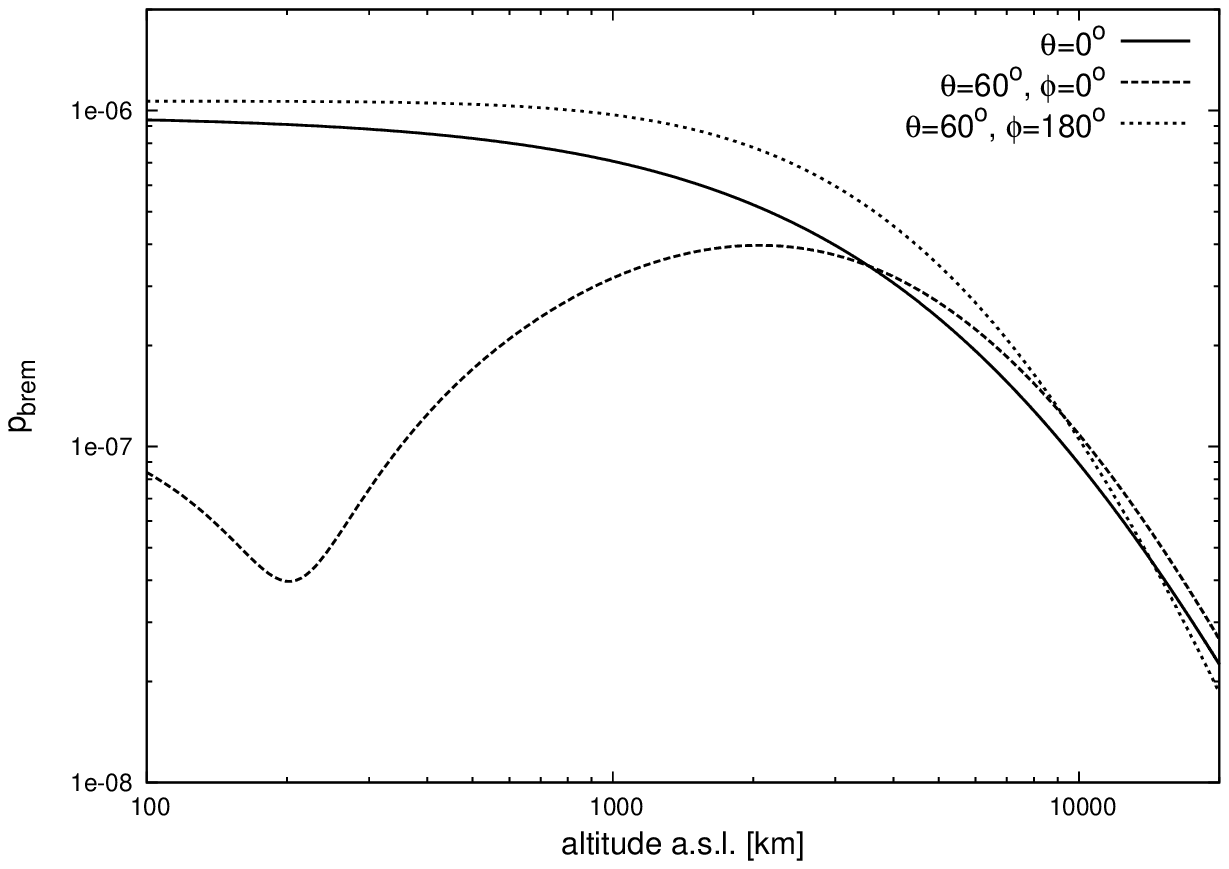}
\end{center}
\caption {
Examples of $p_{brem}$ functions along different trajectories at the location of the Pierre Auger Observatory in  Malarg\"ue (Argentina). A minimum value for one of the curves is related to the small value of $B_\bot$ for this specific arrival direction and altitude. See text for further details.}
\label{alpha_brem}
\end{figure}

The functions $p_{conv/brem}(r)$ depend on $B_\bot(r)$, which is computed with a numerical model. Hence finding the absolute maxima $p^{max}_{conv}$ and $p^{max}_{brem}$ is done numerically. Moreover, through the dependence on $B_\bot(r)$, both $p_{conv}(r)$ and $p_{brem}(r)$ depend on the primary arrival direction and the geographical location of the observatory. As can be seen in Figs.~\ref{alpha_conv} and \ref{alpha_brem}, the values of $p^{max}_{conv}$ and $p^{max}_{brem}$ may be significantly different for various arrival directions.

This indicates that the computation of $p^{max}_{conv}$ and $p^{max}_{brem}$ should be performed for each trajectory separately, otherwise one would have to apply upper limits of these values which would be universal but excessively large for most directions. Accepting the excessive values of $p^{max}_{conv/brem}$ would increase enormously the number of steps in the veto algorithm and might result in an unacceptable increase of the computing time.

Typically, the functions $p_{conv}(r)$ and $p_{brem}(r)$ reach their global maximum at the top of the atmosphere, i.e.\ at the end of preshower simulations, assumed here to be at the altitude of $112$\,km. 
This is the point closest to the Earth's surface and $B_\bot(r)$ typically reaches the maximum value. However for certain classes of trajectories $B_\bot(r)$ might start to decrease with approaching the geomagnetic field source when the trajectory direction approaches a tangent to the local field lines. If this decrease happens to be close to the Earth surface, preshower particles are exposed to the maximum $B_\bot(r)$ somewhere before reaching the atmosphere. Examples of $p^{max}_{conv}$ and $p^{max}_{brem}$ with local extrema are plotted in Figs.~\ref{alpha_conv} and \ref{alpha_brem}. The positions of the minima of $p_{conv/brem}(r)$ are closely correlated with the minima of $B_\bot(r)$.

In case of $p_{conv}(r)$, it has been checked that its global maximum is well reproduced by computing the function value along the trajectory in a simple loop with steps of 1000 km. This procedure is performed for the primary photon energy and the trajectory of interest. It has been checked that $p_{conv}(r)$ decreases with energy, so $p^{max}_{conv}$ found for the primary photon energy will work also for secondary photons of lower energies. In case of electrons, $p_{brem}(r)$ may increase with decreasing electron energy, so one has to compute $p^{max}_{brem}$ for energies within the entire energy range of the simulated particles. Here $p^{max}_{brem}$ is computed in two loops. The external loop along the trajectory is done in steps of 1000 km and the internal loop over energies decreases the energy by one decade in each step. The steps in both procedures can be adjusted by the user if necessary. There is also an alert in the program that gets triggered when the actual values of $p_{conv/brem}(r)$ happen to exceed $p^{max}_{conv/brem}$.

The above method of finding the absolute maxima of $p_{conv/brem}(r)$ is fast and efficient. However, it is optimized only for specific simulation conditions: preshowering in the geomagnetic field. In other environments, involving more irregular shapes of $B(r)$, one has to reconsider the procedure of finding the absolute maxima of $p_{conv/brem}(r)$.

\section{Other modifications and corrections}
\label{sec2}

Other modifications and changes applied in the new release of the PRESHOWER program are briefly listed below.
\begin{enumerate}
 \item The IGRF geomagnetic field model has been updated to the year 2010 and the most recent IGRF-11 coefficients have been applied (Ref.~\cite{igrf11}). In the updated model, the highest order of spherical harmonics has been increased from 10 to 13. The geomagnetic field can be extrapolated up to the year 2015 with the new model. The differences between the field strength and direction in these two models are not larger than 0.001\%.



\item Procedures to calculate modified Bessel functions have been replaced with an open source CERN routine DBSKA.

\item A minor problem has been found in the auxiliary function \texttt{kappa(x)} used for calculation of bremsstrahlung probability. The interpolation performed in this function failed for the rare case of $x=10.0$. This happened because of a faulty definition of the last interval where the interpolation was done. As a result of this bug the input value $x=10.0$ was excluded from the computations. This bug has been fixed in the new release of the program.

\item Since for some cases the number of preshower particles can be very large, the size of the array \texttt{part\_out[50000][8]}, which stores the output particle data, has been increased from 50000 to 100000 entries.

\item The code of the program was reorganized and more clearly structured. The main change here was moving the auxiliary functions and routines to a separate file.

\end{enumerate}
A list of the new and modified files with basic explanations can be found in the Appendix C.

\section{Validation of the new version}
\label{sec3}

The new release has been intensively tested. Below we show some examples to illustrate the performance.

\begin{figure}
\begin{center}
\includegraphics[width=1.0\textwidth]{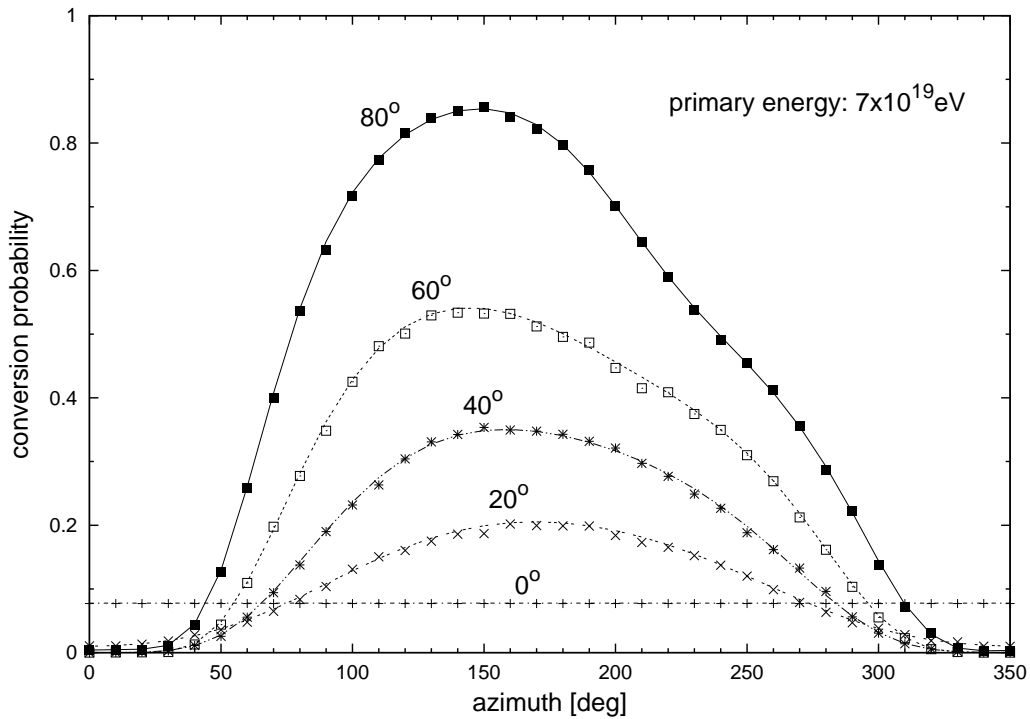}
\end{center}
\caption {
Total probability of $\gamma$ conversion for the primary energy of $7\times 10^{19}$\,eV for different arrival directions as computed by PRESHOWER 1.0 (lines) and PRESHOWER 2.0 (points). PRESHOWER 1.0 values were obtained by numerically integrating the conversion probability in the loop over trajectory. The points represent the fractions of events with gamma conversion simulated by PRESHOWER 2.0 with the new veto algorithm. Each fraction is the average for 10000 primary photons. Computations have been done for magnetic conditions at the Pierre Auger Observatory in Argentina. The azimuth $0^\circ$ refers to showers arriving from the geographic North.}
\label{maps}
\end{figure}

In Fig.~\ref{maps} a comparison of conversion probability obtained with PRESHOWER 1.0 and PRESHOWER 2.0 is presented for different 
arrival directions and primary energy of $7\times 10^{19}$\,eV.
The lines represent conversion probabilities obtained by numerical integrations of the expression (\ref{pconv2}) along trajectories a
nd with steps as given in PRESHOWER~1.0. The points are plotted to show fractions of events with gamma conversion simulated by PRESHOWER~2.0. 
Each fraction was calculated after 10,000 simulation runs. Simulations of gamma conversion probabilities for other primary energies has also
 been checked and in all cases an excellent agreement between the results of the two PRESHOWER versions has been found.

\begin{figure}
\begin{center}
\includegraphics[width=1.0\textwidth,angle=0]{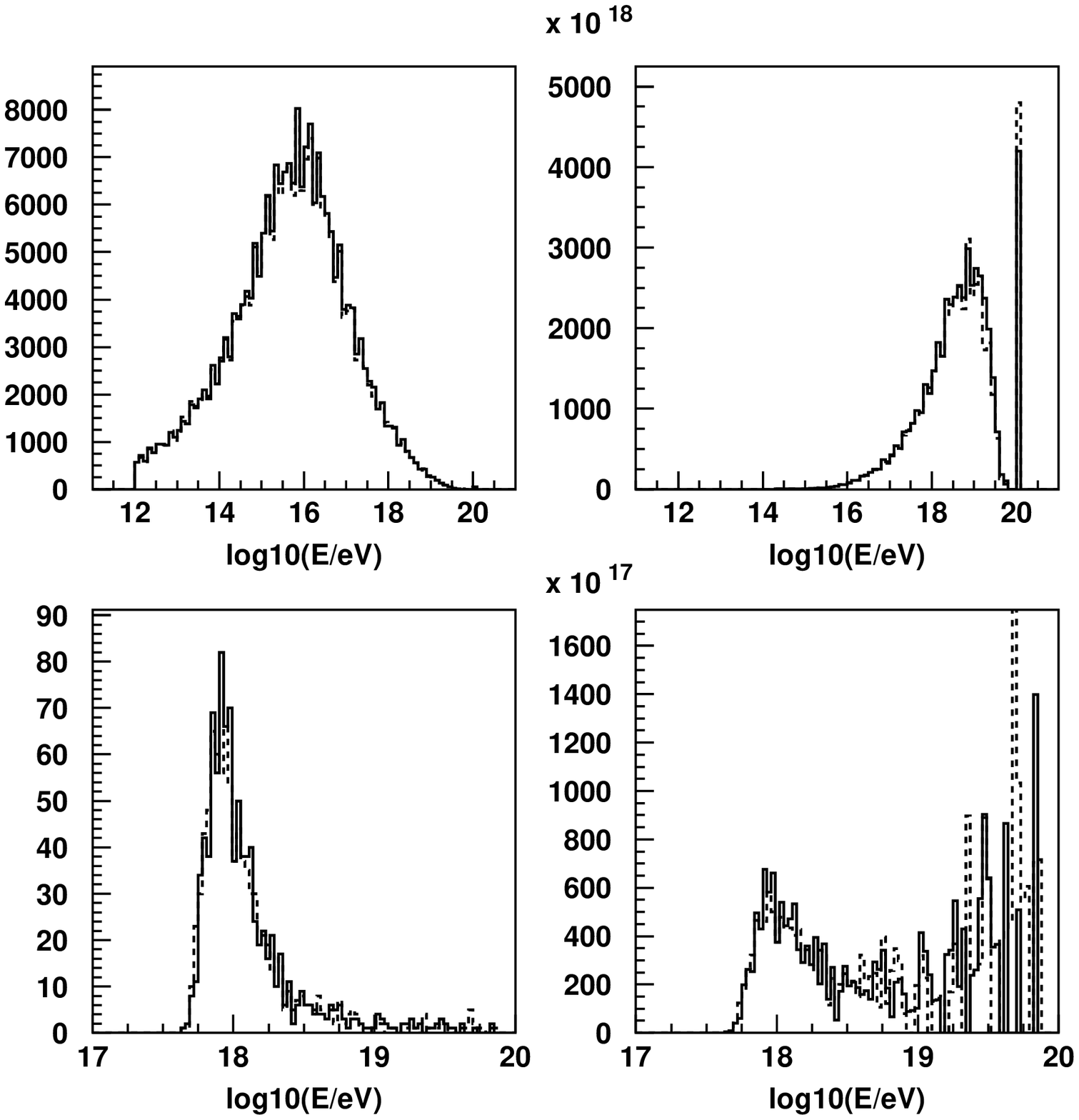}
\end{center}
\caption {Energy distribution of photons (top left) and electrons (bottom left) in 500 preshowers initiated by
$10^{20}$~eV photons arriving at the Pierre Auger Observatory in Argentina from the strong field
direction. The spectra weighted by energy are plotted to the right. The dashed histograms were obtained with PRESHOWER~1.0 and the solid ones 
with PRESHOWER~2.0.}
\label{20strong_profiles}
\end{figure}

A cross-check of the procedures responsible for simulation of bremsstrahlung is shown in Fig.~\ref{20strong_profiles}, in which the energy 
distribution of secondary particles for a primary photon of $10^{20}$\,eV and an arrival direction along which the transverse component of the 
geomagnetic field is particularly strong (``strong field direction'') are compared.
Plotted are the summed distributions of energies of secondary photons and electrons together with the relevant histograms weighted by the energies. 
The summations are done for 500 simulation runs. The results obtained with the two program versions are in very good agreement.

\begin{figure}
\begin{center}
\includegraphics[width=1.0\textwidth,angle=0]{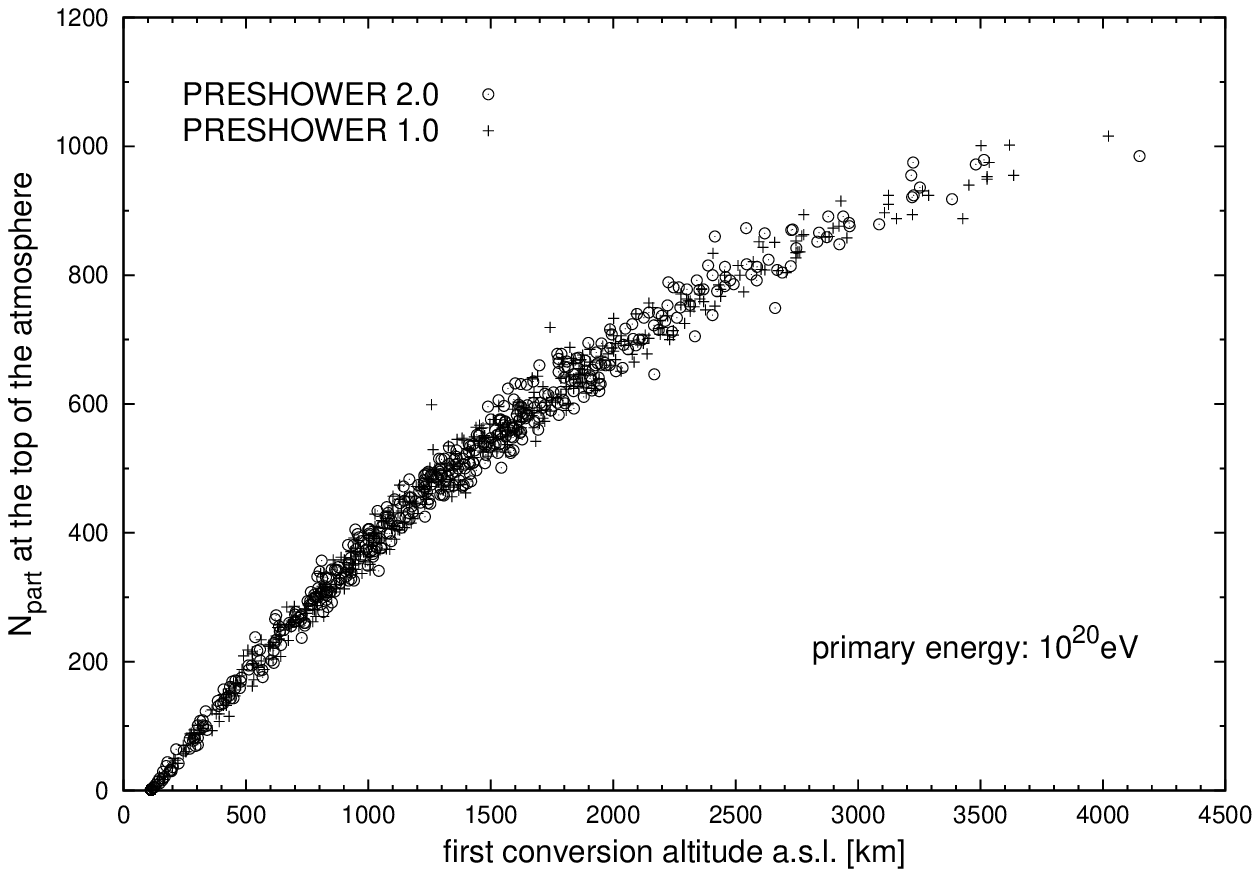}
\end{center}
\caption {Number of particles in the preshower for different altitudes of
the first $\gamma$ conversion simulated with PRESHOWER~1.0 and PRESHOWER~2.0. Plotted are the preshowers initiated by $10^{20}$\,eV photons 
arriving from the strong field direction. The two points  somewhat higher than those of the general trend are cases where one of the bremsstrahlung 
photons again converted in the magnetic field to produce an electron-positron pair which emitted the additional photons. See also Fig.~\ref{cor20efpp}.}
\label{cor20nfpp}
\end{figure}

\begin{figure}
\begin{center}
\includegraphics[width=1.0\textwidth,angle=0]{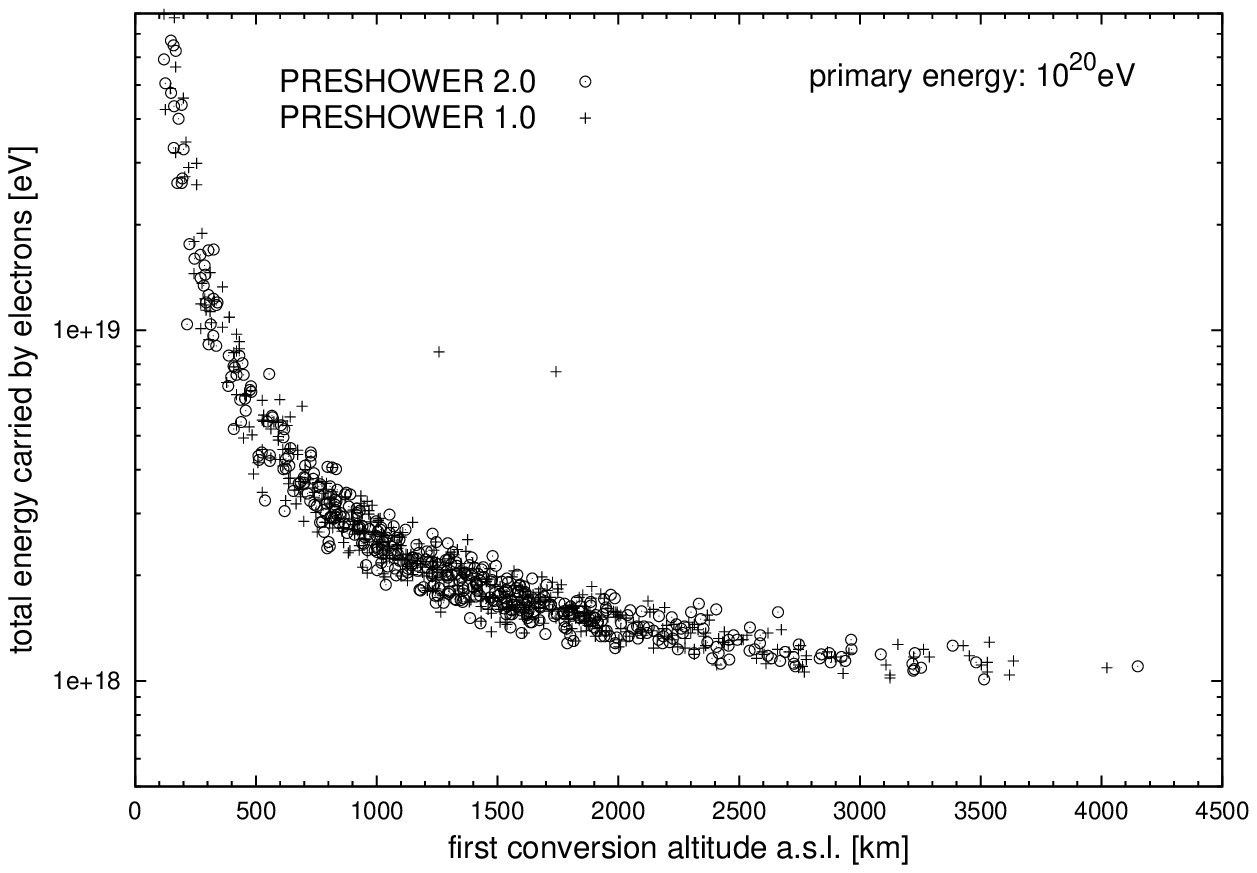}
\end{center}
\caption {Energy carried by preshower electrons at the top of the atmosphere vs.
the altitude of the first $\gamma$ conversion for a primary photon energy of
$10^{20}$~eV in the strong field direction.
The two points in excess of the general trend are two rare cases where the first bremsstrahlung photon converted again into an electron-positron  
pair which increased the total energy carried by leptons. See also Fig.~\ref{cor20nfpp}.}
\label{cor20efpp}
\end{figure}

Further tests for the same set of simulations are shown in Figs.~\ref{cor20nfpp} and \ref{cor20efpp}.
These are the number of preshower particles and the total energy carried by the preshower electrons. Both observables are calculated at the top of 
the atmosphere and both are plotted versus the altitude of primary photon conversion. In both figures a comparison is made between the results obtained 
with PRESHOWER~1.0 and PRESHOWER~2.0. Again, the agreement between the results of the two PRESHOWER versions is very good.

One of the main aims of the new release of PRESHOWER was to reduce the computing time. The results of the CPU time comparison are summarized 
in Table~\ref{table-times}.
\begin{table}
\begin{center}
\caption {The preshower simulation times in PRESHOWER 1.0 (old) and PRESHOWER 2.0 (new) for selected arrival directions and primary energies. The arrival directions are selected to represent a typical variation of $B_\bot$:
$\theta=70^o$ and $\phi=0^o$ for a ``weak'' $B_\bot$, $\theta=0^o$ for a ``medium'' $B_\bot$ and $\theta=60^o$ and $\phi=177^o$ for a ``strong'' $B_\bot$.}
\vspace{0.5cm}
\begin{tabular}
{rcp{2.0cm}p{1.7cm}p{1.7cm}} \hline \hline
$E_0$ [eV]& direction & fraction of converted & time old [sec.] & time new [sec.] \\ \hline
7$\times$10$^{19}$ & $\theta=0^o$ & 0/1000 & 79 & 7 \\
7$\times$10$^{19}$ & $\theta=70^o$, $\phi=0^o$ & 0/1000 & 76 & 8 \\
10$^{20}$ & $\theta=60^o$, $\phi=177^o$  & 92/100 & 1195 & 209 \\ \hline \hline
\end{tabular}
\label{table-times}
\end{center}
\end{table}
The computing time is more than a factor 5 shorter in case of simulations with PRESHOWER~2.0 than in the case of PRESHOWER~1.0. 
This reduction is seen both in computation of gamma conversion (speed up by nearly factor 10) and in more time consuming bremsstrahlung routines.

\section{Summary}
\label{sec4}

The program PRESHOWER is a tool designed for simulating magnetically induced particle cascades due to ultra-high energy photons. It can be linked with air 
shower simulation packages such as CORSIKA \cite{corsika} to calculate complete photon-induced particle cascades as they are searched for with cosmic ray 
observatories.

A new version of the PRESHOWER program, version 2.0, has been released and its features are presented in this article. An efficient veto algorithm has been 
introduced to sample the locations of individual interaction processes. Other modifications include the update of the geomagnetic field model, correcting a 
rare exception, and reorganizing the program code. The results obtained with the new release of PRESHOWER agree very well with those calculated with the 
previous version.

The new algorithm not only speeds up the program by more than a factor 5, but also allows additional applications due to the increased flexibility of the 
sampling of interaction points. For example, the preshower effect can now be studied not only in the geomagnetic field, but also close to extended
 astrophysical objects like neutron stars and active galactic nuclei. An application of PRESHOWER 2.0 in the conditions other than the geomagnetic 
field require only small changes in the program. The magnetic field model has to be replaced and start and end points of simulations have to be 
adequately adjusted.

\section*{Acknowledgements}

We thank N.A.~Tsyganenko for valuable remarks on the application of the IGRF model. We are also thankful to Carla Bleve whose update of the IGRF coefficients in Tsyganenko's subroutine has been used.\\

This work was partially supported by the Polish Ministry of Science and Higher Education under grant No. N~N202~2072~38 and by the DAAD (Germany) under grant No.~50725595.

\appendix

\section{Magnetic pair production: $\gamma \rightarrow e^+e^-$}
\label{magneticpp}
The number of pairs created by a high-energy photon in the presence of a magnetic
field per path length $dr$
can be expressed in terms of the attenuation coefficient $\alpha(\chi)$ \cite{Erber}:
\begin{equation}
\label{npairs}
n_{pairs}=n_{photons}\{1-\exp[-\alpha(\chi)dr]\},
\end{equation}
where
\begin{equation}
\label{alpha}
\alpha(\chi)=0.5(\alpha_{em} m_ec/\hbar)(B_\bot/B_{cr})T(\chi)
\end{equation}
with $\alpha_{em}$ being the fine structure constant, $\chi\equiv0.5(h\nu/m_ec^2)(B_\bot/B_{cr})$,
$B_\bot$ is the magnetic field component transverse to the direction of the photon's motion,
$B_{cr}\equiv m_e^2c^3/e\hbar=4.414\times 10^{13}$ G
and $T(\chi)$ is the magnetic pair production function. $T(\chi)$ can be well approximated by:
\begin{equation}
\label{tcentral}
T(\chi)\cong0.16\chi^{-1}{K^2}_{1/3}(\frac{2}{3\chi}),
\end{equation}
where $K_{1/3}$ is the modified Bessel function of order $1/3$. For small or large arguments
$T(\chi)$ can be approximated by

\begin{equation}
\label{tlimits}
\begin{array}{c}
T(\chi)\cong\left\{ \begin{array}{ll}
0.46\exp(-\frac{4}{3\chi}), & ~~\chi \ll 1;\\
0.60\chi^{-1/3}, & ~~\chi \gg 1.
\end{array} \right.
\end{array}
\end{equation}

We use Eq.~(\ref{npairs})
to calculate the probability of $\gamma$ conversion over a small path length $dr$:
\begin{equation}
\label{pconv2}
p_{conv}(r)=1-\exp[-\alpha(\chi(r))dr]\simeq\alpha(\chi(r))dr.
\end{equation}

\section{Magnetic bremsstrahlung}
\label{a2}
After photon conversion, the electron-positron pair is propagated.
The energy distribution in an $e^+e^-$ pair
is computed according to Ref. \cite{ppdaugherty}:

\begin{equation}
\frac{d\alpha(\varepsilon,\chi)}{d\varepsilon}\approx\frac{\alpha_{em}m_ec B_\bot}{\hbar B_{cr}}
\frac{3^{1/2}}{9\pi\chi}\frac{[2+\varepsilon(1-\varepsilon)]}{\varepsilon(1-\varepsilon)}
K_{2/3}\left[\frac{1}{3\chi\varepsilon(1-\varepsilon)}\right],
\label{daug}
\end{equation}
where $\varepsilon$ denotes the fractional energy of an electron and the other symbols
were explained in the previous chapter.
The probability of asymmetric energy partition grows with the primary photon energy and
with the magnetic field. Beginning from $\chi>10$, the asymmetric energy partition is
even more favored than the symmetric one.

Electrons traveling at relativistic speeds in the presence of a magnetic field
emit bremsstrahlung radiation (synchrotron radiation). For electron energies
$E \gg m_ec^2$ and for $B_\bot \ll B_{cr}$, the spectral distribution of radiated
energy is given in Ref. \cite{sokolov}:
\begin{equation}
f(y)=\frac{9\sqrt{3}}{8\pi}\frac{y}{(1+\xi y)^3}\left\{\int^\infty_yK_{5/3}(z)dz+
\frac{(\xi y)^2}{1+\xi y}K_{2/3}(y)\right\},
\label{fy}
\end{equation}
where $\xi =(3/2)(B_\bot/B_{cr})(E/m_ec^2)$, $E$ and $m_e$ are electron initial energy
and rest mass respectively, $K_{5/3}$ and $K_{2/3}$ are modified Bessel functions,
and $y$ is related to the emitted photon energy $h\nu$ by
\begin{equation}
y(h\nu)=\frac{h\nu}{\xi (E-h\nu)} \;; \qquad \qquad
\frac{dy}{d(h\nu)}=\frac{E}{\xi(E-h\nu)^2}.
\label{yhv}
\end{equation}
The total energy emitted per unit distance is (in CGS units)
\begin{equation}
W=\frac{2}{3}r_0^2B_\bot^2\left(\frac{E}{m_ec^2}\right)^2\int^\infty_0f(y)dy
\label{W}
\end{equation}
with $r_0$ being the classical electron radius. For our purposes we use the spectral distribution
of radiated energy defined as
\begin{equation}
I(B_\bot,E,h\nu)\equiv\frac{h\nu dN}{d(h\nu)dx}~~,
\label{Idef}
\end{equation}
where $dN$ is the number of photons with energy between $h\nu$ and $h\nu+d(h\nu)$
emitted over a distance $dx$. From Eqs.~(\ref{fy}), (\ref{yhv}), (\ref{W}),
and (\ref{Idef})
we get\footnote{
Expression (\ref{brem}), valid for all values of $h\nu$, is equivalent to Eq. (2.5a)
in Ref. \cite{Erber}. A simplified form of distribution (\ref{brem}) is given by Eq. (2.10)
in Ref. \cite{Erber}, however it can be used only for $h\nu \ll E$.}
\begin{equation}
I(B_\bot,E,h\nu)=\frac{2}{3}r_0^2B_\bot^2\left(\frac{E}{m_ec^2}\right)^2f(h\nu)\frac{E}
{\xi(E-h\nu)^2}~~.
\label{brem}
\end{equation}
Provided $dx$ is small enough, $dN$ can be interpreted as a probability
of emitting a photon of energy between $h\nu$ and $h\nu+d(h\nu)$ by an electron
of energy $E$ over a distance $dx$. In our simulations we use a small step size of $dx=1$ km.
The total probability of emitting a photon in step $dx$ can then be written as
\begin{equation}
P_{brem}(B_\bot,E,h\nu,dx)=\int dN=dx\int^E_0 I(B_\bot,E,h\nu)\frac{d(h\nu)}{h\nu}~~.
\label{bremprob}
\end{equation}
The energy of the emitted photon is simulated according to the
probability density distribution $dN/d(h\nu)$ obtained from Eq.~\ref{bremprob}.

\section{Details on the new and modified files included in the PRESHOWER 2.0 package}
The files included in PRESHOWER 2.0 package but not existing in the previous release are :
\begin{description}
\item[\texttt{IGRF-11.f}] An external routine generating the geomagnetic field components according to the IGRF-11 model \cite{tsygan}. 
This file replaces \texttt{igrf.f} of the previous release of PRESHOWER.
\item[\texttt{cernbess.f}] External procedures calculating the sequence of modified Bessel functions \cite{cernbess}. These open source procedures 
replace previously used functions from Numerical Recipes.
\item[\texttt{veto.c}] This file contains functions and procedures called by the veto algorithm.
\item[\texttt{veto.h}] The header file for \texttt{veto.c}.
\item[\texttt{utils.c}] This file contains auxiliary functions and procedures used within the program. In the previous version of 
PRESHOWER the auxiliary functions were placed in \texttt{preshw.c}, now it is more convenient to have them in a separate file.
\item[\texttt{utils.h}] The header file for \texttt{utils.c}.\\
\end{description}

The list of files which existed in the previous release of PRESHOWER but have been modified for PRESHOWER 2.0 include:
\begin{description}
\item[\texttt{preshw.c}] contains the main procedure \texttt{preshw\_veto} generating preshowers with the veto algorithm,
\item[\texttt{prog.c}] reads input parameters and calls \texttt{preshw\_veto},
\item[\texttt{Makefile}] modified to account for the new files.
\end{description}







\end{document}